\documentclass[aps,reprint,superscriptaddress,amsmath,showpacs,byrevtex]{revtex4-1}
\usepackage{graphicx}
\usepackage{units}
\usepackage{multirow}
\usepackage{color}
\newcommand{\jpsi}{J/\psi}
\newcommand{\etac}{\eta_{c}}
\newcommand{\etap}{\eta^{\prime}}
\newcommand{\psip}{\psi(3686)}
\newcommand{\gm}{\gamma}
\newcommand{\mev}{\,\unit{MeV}}

\newcommand{\gev}{\,\unit{GeV}}
\newcommand{\gevc}{\,\unit{GeV}/\unit{c}}
\newcommand{\gevcc}{\,\unit{GeV}/\unit{c}^2}
\newcommand{\br}[1]{\mathcal{B}(#1)}
\newcommand{\chisq}[1]{\chi^{2}_{\unit{#1}}}

\begin{document}%%

\title{\boldmath Evidence for $\etac\to\gm\gm$ and Measurement of $\jpsi\to3\gm$}

\author{
\small
\begin{center}
M.~Ablikim$^{1}$, M.~N.~Achasov$^{6}$, D.~J.~Ambrose$^{39}$, F.~F.~An$^{1}$, Q.~An$^{40}$, Z.~H.~An$^{1}$, J.~Z.~Bai$^{1}$, Y.~Ban$^{26}$, J.~Becker$^{2}$, J.~V.~Bennett$^{16}$
% ...
\\
\vspace{0.2cm}
(BESIII Collaboration)\\
\vspace{0.2cm} {\it
$^{1}$ Institute of High Energy Physics, Beijing 100049, People's Republic of China\\
$^{2}$ Bochum Ruhr-University, 44780 Bochum, Germany\\
% ...
$^{6}$ G.I. Budker Institute of Nuclear Physics SB RAS (BINP), Novosibirsk 630090, Russia\\
% ...
$^{16}$ Indiana University, Bloomington, Indiana 47405, USA\\
% ...
$^{26}$ Peking University, Beijing 100871, People's Republic of China\\
% ...
$^{39}$ University of Rochester, Rochester, New York 14627, USA\\
$^{40}$ University of Science and Technology of China, Hefei 230026, People's Republic of China\\
% ...
}\end{center}
\vspace{0.4cm}
}
%%%%%%%%%%%%%%%%%%%%%%%%%%%%%%%%%%%%%%%%%%%%%%%%%%%%%%%%%%%%%%%%%%%%%%%%%%%%%%%%%%%%%%%%%%

\begin{abstract}

The decay of $\jpsi$ to three photons is studied using
$\psip\to\pi^+\pi^-\jpsi$ in a sample of $1.0641\times10^8$ $\psip$ events
collected with the BESIII detector.  Evidence of the direct decay of $\etac$ to two photons,
$\etac\to\gm\gm$, is reported, and the product branching fraction is
determined to be $\br{\jpsi\to\gm\etac,\etac\to
\gm\gm}=(4.5\pm1.2\pm0.6)\times10^{-6}$, where the first error is
statistical and the second is systematic.  The branching fraction for $\jpsi\to3\gm$ is measured to be $(11.3\pm1.8\pm2.0)\times 10^{-6}$ with improved precision.

\end{abstract}


\maketitle

%%%%%%%%%%%%%%%%%%%%%%%%%%%%%%%%%%%%%%%%%%%%%%%%%%%%%%%%%%%%%%%%
%%%%%     Introduction       Part                  %%%%%%%%%%%%%
%%%%%%%%%%%%%%%%%%%%%%%%%%%%%%%%%%%%%%%%%%%%%%%%%%%%%%%%%%%%%%%%

Decays of positronium to more than one photon are regarded as an ideal
test-bed for quantum electrodynamics (QED)~\cite{Karshenboim:2003vs}, while the analogous processes in
charmonia act as a probe of the strong
interaction~\cite{Czarnecki:2001zc}.  For example, the decay
$\jpsi\to3\gm$ has a relatively simple theoretical
description, and the experimental measurements allow for a fundamental test of
non-perturbative quantum chromodynamics (QCD)~\cite{Hagiwara:1980nv}.
The decay rate of $\jpsi\to3\gm$ is approximately proportional to the cube of the QED coupling constant $\alpha^3\approx(\frac{1}{137})^3$.
To reduce model dependence, the branching fraction for $\jpsi\to3\gm$ is normalized by the branching fraction for
$\jpsi\to e^+e^-$. The ratio
\begin{equation}
\mathcal{R}\equiv\frac{\br{\jpsi\to3\gm}}{\br{\jpsi\to e^+e^-}}=\frac{64(\pi^2-9)}{243\pi}\alpha(1-7.3\frac{\alpha_s(r)}{\pi})
\label{eq:jpsi_ratio}
\end{equation}
is calculated with first-order QCD corrections, where
$\br{\unit{X}}$ denotes the branching fraction of decay
X, $\alpha_s(r)$ is the QCD running coupling constant, and $r$ is the distance between the $c$ and $\bar{c}$ quarks.
From the ratio $\br{\jpsi\to3g}/\br{\jpsi\to e^+e^-}$~\cite{Besson:2008pr}, a value of $\alpha_s\approx0.19$ can be obtained; inserting this into Eq.~\eqref{eq:jpsi_ratio} then gives $\mathcal{R}\approx2.96\times10^{-4}$.
%For $\alpha_s\approx0.19$, which is extracted from the ratio of $\br{\jpsi\to3g}/\br{\jpsi\to e^+e^-}$~\cite{Besson:2008pr}, Eq.~\eqref{eq:jpsi_ratio} predicts $\mathcal{R}\approx2.96\times10^{-4}$.
This ratio is sensitive to QCD corrections only. It is still unclear, though, how radiative and relativistic QCD corrections should be treated~\cite{Mackenzie:1981sf} and how they may affect this ratio.
Experimental constraints on this ratio can help us to understand the behavior of non-perturbative QCD, which would shed light on the dynamics of charmonium. In addition, the photon energy spectrum in $\jpsi\to3\gm$ reveals the internal structure of the $\jpsi$, since the photon spectrum at energy $\omega$ is sensitive to the distance $r \sim 1/\sqrt{m_c\omega}$~\cite{Voloshin:2007dx}.

The CLEO collaboration was the first to report the observation of $\jpsi\to3\gm$,
measuring its branching fraction to be $\br{\jpsi\to3\gm}=(12\pm3\pm2)\times10^{-6}$~\cite{Adams:2008ab}. This corresponds to a value of $\mathcal{R}=(2.0\pm0.6)\times10^{-4}$, which disagrees with the prediction given by Eq.~\eqref{eq:jpsi_ratio}.
Looking at the $\jpsi\to\gm\etac, \etac\to\gm\gm$ mode, the analysis of $\br{\etac\to\gm\gm}$ is determined mainly from
two-photon fusion $\gm\gm^{(*)}\to\etac$~\cite{PDG2012}, because of low statistics for
direct measurements of the decay.  The most precise direct
measurement of $\br{\etac\to\gamma\gamma}$ to date comes from BELLE, with a significance of $4.1\sigma$~\cite{Abe:2006gn}.
The $\jpsi\to\gm\etac, \etac\to\gm\gm$ branching fraction is
predicted to be $(4.4\pm1.1)\times10^{-6}$~\cite{Kwong:1987ak}, if higher-order QCD corrections are not taken into account.
CLEO reported an upper limit of $\br{\jpsi\to\gm\etac, \etac\to\gm\gm} <6\times10^{-6}$ at 90\% confidence level~\cite{Adams:2008ab}.

This article presents the most precise measurement yet of the $J/\psi\to3\gamma$ branching fraction and its photon energy spectrum using $\psip\to\pi^+\pi^-\jpsi$ decays. In addition, evidence for $\jpsi\to\gm\etac, \etac\to\gm\gm$ is reported. The analysis is based on a sample of $(1.0641\pm0.0086)\times10^8$ $\psip$ events~\cite{Ablikim:2012pj}
collected with the Beijing Spectrometer (BESIII), at the
Beijing Electron-Positron Collider (BEPCII)~\cite{:2009vd}.
Using $\psip\to\pi^+\pi^-\jpsi$ events for this study rather than $e^+e^- \to
\jpsi \to 3 \gamma$ eliminates background from the QED process
$e^+e^- \to 3 \gamma$.

BEPCII is a double-ring electron-positron collider, designed to run at energies around the $\jpsi$ peak.  The
BESIII detector~\cite{:2009vd} is a cylindrically symmetric detector
with five sub-detector components. From inside to out, these are: main
drift chamber (MDC), time-of-flight system,
electromagnetic calorimeter (EMC), super-conducting solenoid
magnet, and muon chamber.  The momentum resolution for charged
tracks reconstructed by the MDC is $0.5\%$ for transverse momenta of
$1\gevc$.  The energy resolution for showers deposited in the EMC is
$2.5\%$ for $1\gev$ photons.

The BESIII detector is modeled with a Monte Carlo (MC) simulation
based on GEANT4~\cite{ref:geant4,bib:boost}.
The KKMC generator~\cite{Jadach:2000ir} is used to produce MC samples at any
specified energy, taking into account initial state radiation and beam energy spread.  The known $\psip$ decay modes are generated with
EVTGEN~\cite{ref:bes3gen} using branching fractions listed by the Particle Data
Group (PDG)~\cite{PDG2012}, while unknown decay modes are simulated with LundCharm~\cite{Chen:2000tv}.

For the selection of $\psip\to\pi^+\pi^-\jpsi$, $\jpsi\to3\gm$
candidates, events with only two charged tracks and at least three
photons are required.  The minimum distance of any charged track to the
interaction point is required to be within 10\,cm in the beam
direction and within 1\,cm in the perpendicular plane. The two charged
tracks are assumed to be $\pi^+\pi^-$ candidates, and the recoil mass
in the center of mass system must be in the range [$3.091, 3.103$]$\gevcc$.

Photon candidates are chosen from isolated clusters in the EMC whose
energies are larger than 25\,MeV in the barrel region
($|\cos\theta|<0.8$) and 50\,MeV in the end-cap regions
($0.86<|\cos\theta|<0.92$). Here, $\theta$ is the polar angle with
respect to the beam direction. To reject photons from bremsstrahlung
and from interactions with material, showers within a conic angle of
$5^\circ$ around the momenta of charged tracks are rejected.
%CN - around the momenta of charged tracks??
To suppress wrongly reconstructed showers due to electronic noise or beam backgrounds, it is required that the shower time be within 700\,ns of the event start time.
%Reconstructed showers due to electronic noise or beam backgrounds are
%suppressed by limiting the timing information to within [0, 700]\,ns
%after the event start time.
Events with 3 or 4 photon candidates
are kept for further data processing.

The $\pi^+$ and $\pi^-$ tracks are fitted to a common vertex to determine
the event interaction point, and a four-constraint kinematic fit to the initial
four-momentum of the $\psip$ is applied for each $\pi^+\pi^-\gm\gm\gm$
combination.  The combination with the smallest fit
$\chisq{4C}$ is kept, and $\chisq{4C}<50$ is required.

%%%%%%%%%%%%%%%%%%%%%%%%%%%%%
\begin{figure}[t]
    \centering
    \includegraphics[width=\linewidth]{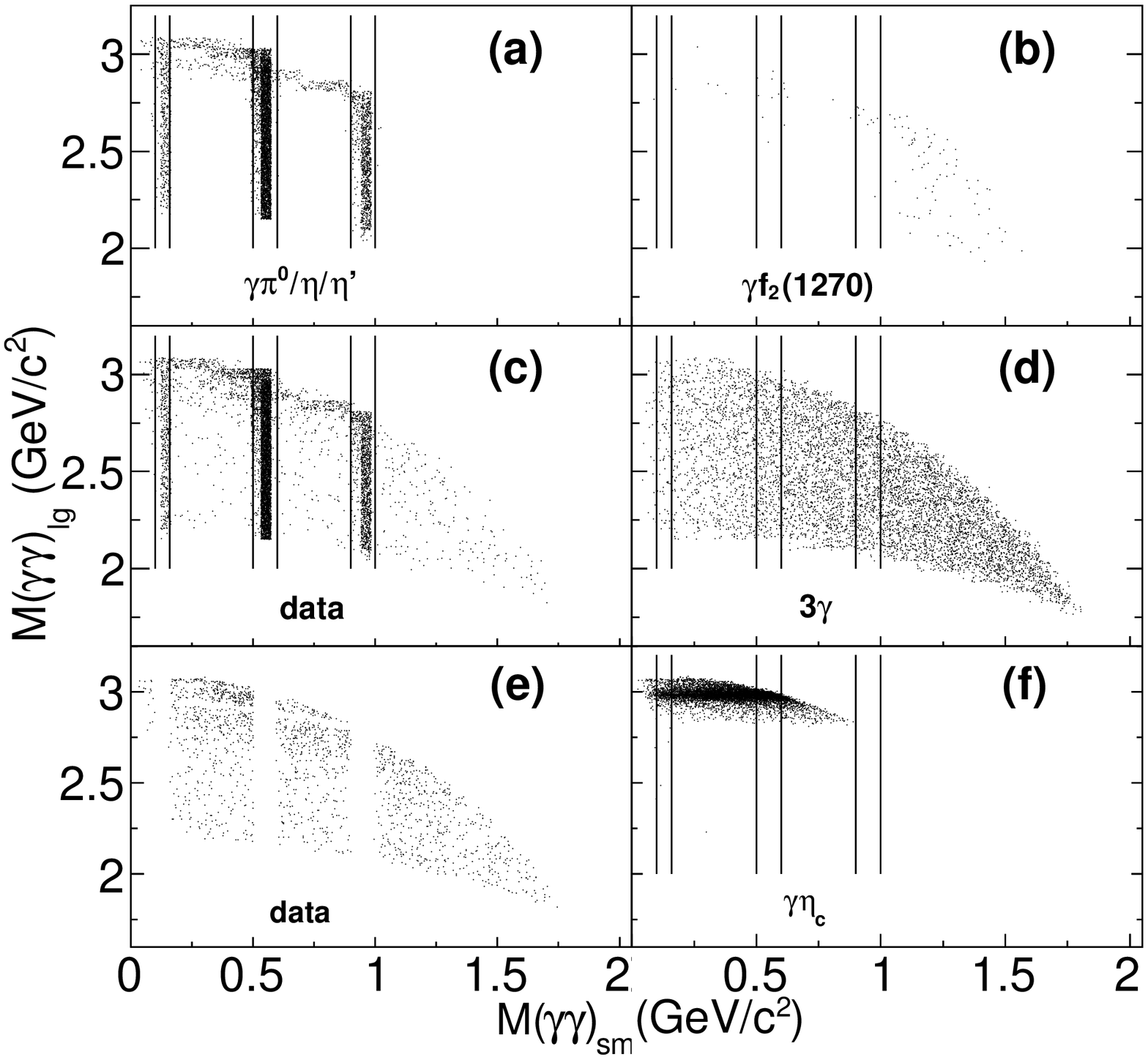}
    \vspace{-0.25cm}
    \caption{Scatter plots of $M(\gm\gm)_{\unit{lg}}$ versus
$M(\gm\gm)_{\unit{sm}}$ for data before (c) and after (e) removal of backgrounds from $\jpsi\to\gm\pi^0/\eta/\eta'$ and MC simulations of the
processes (a) $\jpsi\to\gm\pi^0/\eta/\eta'\to3\gm$, (b) $\jpsi\to\gm
f_2(1270)\to\gm(\gm\gm)_{\pi^0}(\gm\gm)_{\pi^0}$, (d) $\jpsi\to3\gm$,
and (f) $\jpsi\to\gm\etac\to3\gm$. The vertical lines
indicate the mass windows to reject $\pi^0$, $\eta$ and
$\eta^\prime$.}
    \vspace{-0.25cm}
    \label{fig:proj}
\end{figure}
%%%%%%%%%%%%%%%%%%%%%%%%%%%%%%%%%%%%%%%%%%%%%%%%%%%%%%%%%%%%%%%%%%%%%%

Figure~\ref{fig:proj} shows distributions of $M(\gm\gm)_{\unit{lg}}$ versus
$M(\gm\gm)_{\unit{sm}}$, where
$M(\gm\gm)_{\unit{lg}}$ and $M(\gm\gm)_{\unit{sm}}$ are the largest
and smallest two-photon invariant masses among the three combinations,
respectively. Events from the background processes
$\jpsi\to\gm\pi^0/\eta/\eta'\to3\gm$ can be clearly seen in Fig.~\ref{fig:proj}(c). These backgrounds
are significantly reduced by removing all events that lie in the mass regions
$[0.10,0.16]\gevcc$, $[0.50, 0.60]\gevcc$, and $[0.90,1.00]\gevcc$. Contributions
from these backgrounds which lie outside these mass regions are estimated from simulation.
Simulations of these processes are validated by comparing
the line shapes of the $M(\gm\gm)_{\unit{lg}}$ and $M(\gm\gm)_{\unit{sm}}$ distributions and their yields with those in the control samples in data.

Another source of background is $\jpsi\to\gm e^+e^-$ events in which the electron and positron tracks fail to be reconstructed in the MDC, with the associated EMC clusters then being misidentified as photon candidates.
To reject this background, the number of hits in the MDC
within an opening angle of five EMC crystals around the center
of each photon shower is counted and the total number of hits
from the three photons is required to be less than 40.

\begin{figure}[t]
\centering
\includegraphics[width=1.0\linewidth]{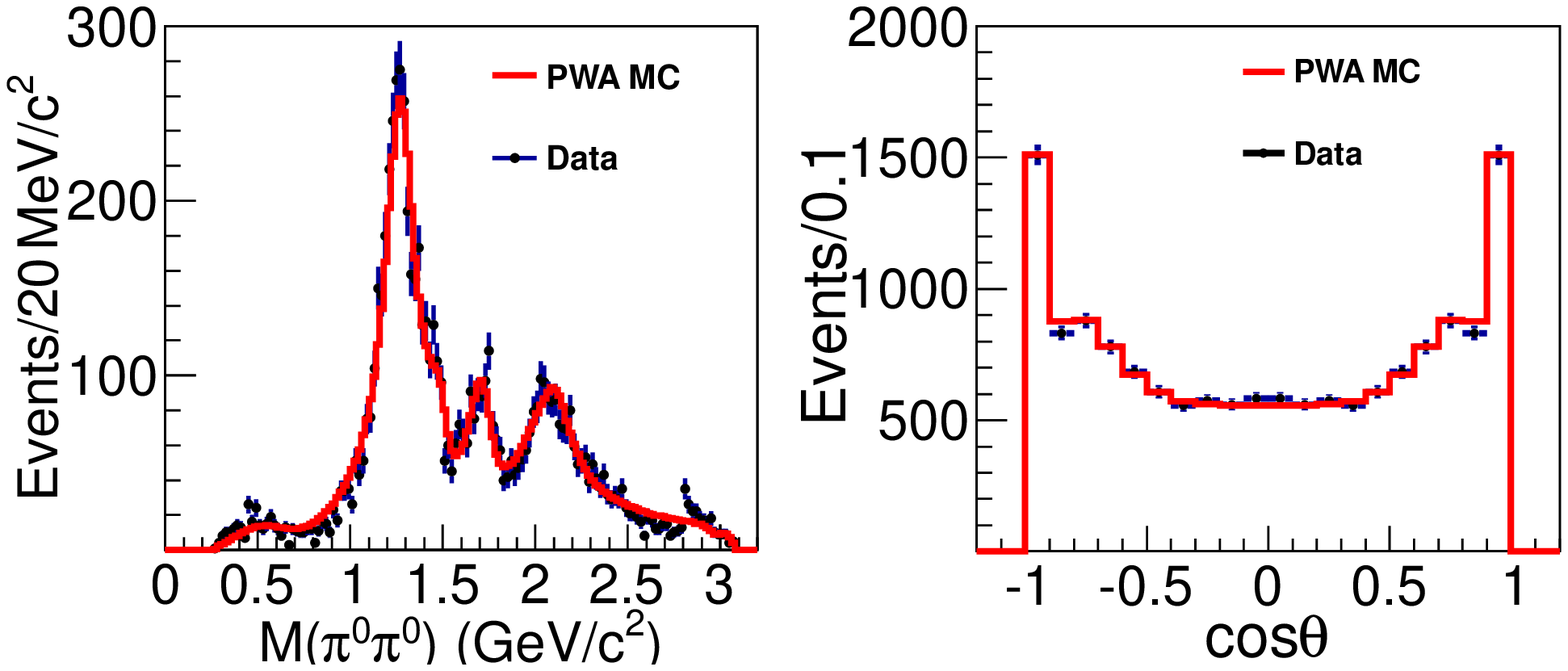}
\vspace{-0.25cm}
\caption{The $\pi^0\pi^0$ invariant mass spectrum (left) and
the angular distribution of the $\pi^0$ in the laboratory frame (right)
for the $\psip\to\pi^+\pi^-\jpsi$, $\jpsi\to\gm\pi^0\pi^0$ control sample, for data (points with error bars) and PWA results (solid line).}
\vspace{-0.25cm}
\label{fig:dect:gpp}
\end{figure}

Background from $\jpsi\to\gm{\pi^0}{\pi^0}$ events can still pass the selection requirements
if the two photons from one $\pi^0$ decay are nearly collinear
or if one of the $\pi^0$s is very soft. Since the $\jpsi\to\gm\pi^0\pi^0$ branching fraction
is large, this remains a large source of
background. In order to model this background, taking advantage of the structure of
intermediate resonances, a partial wave analysis (PWA)~\cite{Ablikim:2006db}
is performed on a $\gm\pi^0\pi^0$ sample based on $2.25\times10^8$ $\jpsi$ events recorded at the $\jpsi$ resonance at BESIII~\cite{liubj}.
The intermediate states $f_0(600)$, $f_2(1270)$, $f_0(1500)$,
$f^\prime_2(1525)$, $f_0(1710)$, $f_2(1950)$,
$f_0(2020)$, $f_2(2150)$ and $f_2(2340)$ are probed and measured in the $\gm\pi^0\pi^0$ final states of $\jpsi$ decays.
For the control samples of $\jpsi\to\gm{\pi^0}{\pi^0}$ in $\psip\to\pi^+\pi^-\jpsi$ decays, looking at the distributions of $M(\pi^0\pi^0)$ and $\cos\theta$, Fig.~\ref{fig:dect:gpp} shows excellent agreement between
data and MC simulation which incorporates the PWA results.
Here, $M(\pi^0\pi^0)$ is the invariant mass of two $\pi^0$ and $\theta$ is the polar angle of the $\pi^0$
with respect to the beam axis.
Decays of $\jpsi\to\gm f_{J}$, $f_J\to\gm\gm$ are negligible because
of their extremely small branching fractions~\cite{PDG2012}.

\begin{figure}[t]
\centering
\includegraphics[width=\linewidth]{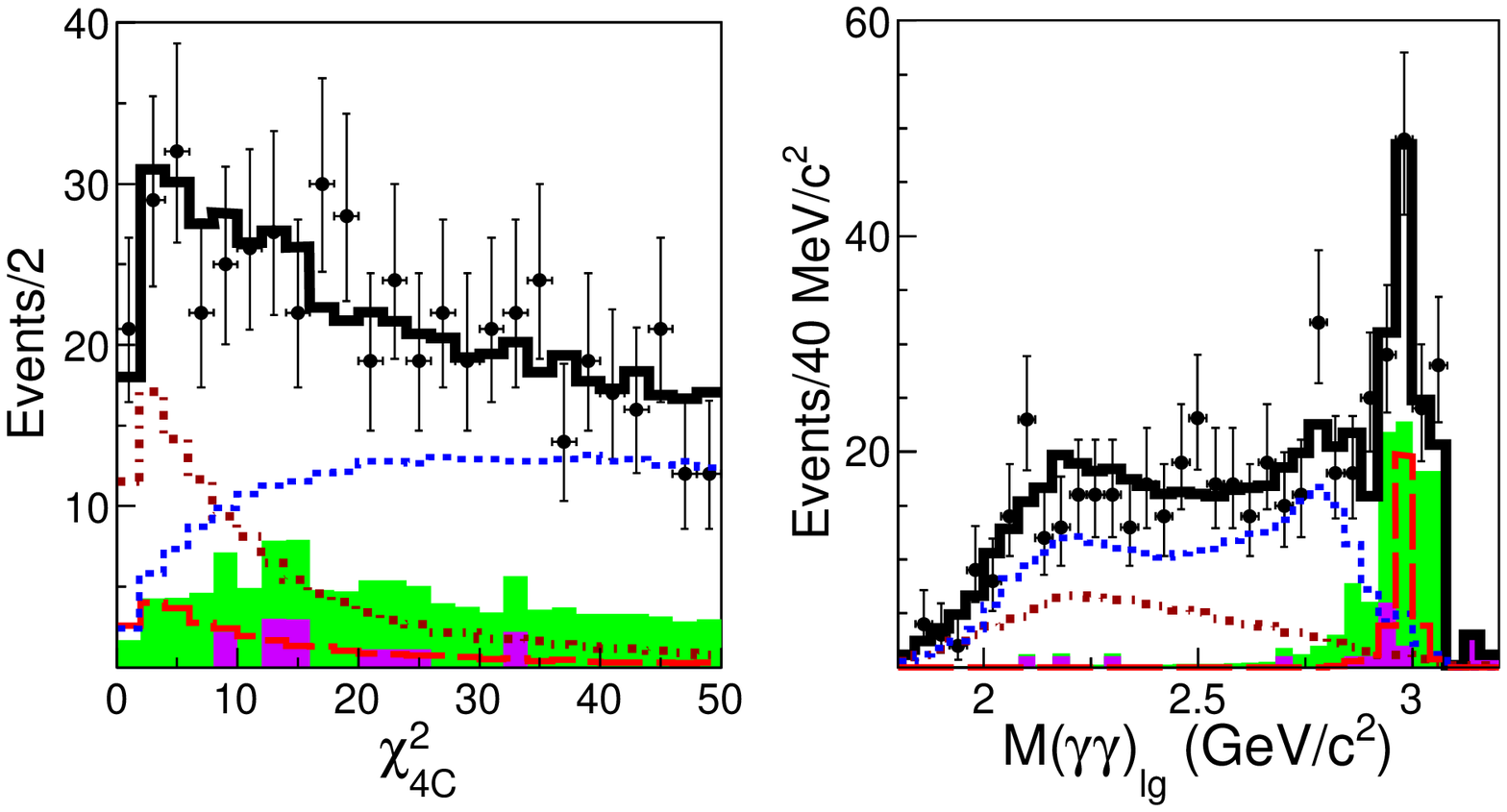}
\vspace{-0.3cm}
\caption{(color online) Projection of the two-dimensional fit to
$\chisq{4C}$ (left) and $M(\gm\gm)_{\unit{lg}}$ (right) for data (points
with error bars) and the fit results (thick solid line).
The (dark red) dotted-dashed, (red) dashed and (blue) dotted lines show
contributions from $\jpsi\to3\gm$, $\jpsi\to\gm\etac\to3\gm$, and
$\jpsi\to\gm\pi^0\pi^0$, respectively. The stacked histogram represents the
backgrounds from $\jpsi\to\gm\pi^0/\eta/\eta'$ (light shaded and green) and non-$\jpsi$ decays (dark shaded and violet).}
\vspace{-0.25cm}
\label{fig:fitting2d}
\end{figure}

The $\chisq{4C}$ value can be used to separate the $3\gm$ from
the $\gm\pi^0\pi^0$ final states, and the $M(\gm\gm)_{\unit{lg}}$
distribution can be used to distinguish $\jpsi\to\gm(\gm\gm)_{\etac}$
from the direct process $\jpsi\to3\gm$. A two-dimensional
maximum likelihood fit is therefore performed on the $M(\gm\gm)_{\unit{lg}}$ and
$\chisq{4C}$ distributions to estimate the yields of
$\jpsi\to3\gm$ and $\jpsi\to\gm(\gm\gm)_{\etac}$.
For the fit, the shapes of both signal and background processes are
taken from MC simulation; the normalization of
$\jpsi\to\gm(\gm\gm)_{\pi^0/\eta/\eta'}$ is fixed to the expected
density based on MC simulation as listed in Table~\ref{tbl:num_bkg};
and the normalization of
$\jpsi\to\gm\pi^0\pi^0$ is allowed to float.  Backgrounds of
non-$\jpsi$ decays are estimated using the
$M(\pi^+\pi^-)_{\unit{recoil}}$ sidebands within [$2.994,
3.000$]$\gevcc$ and [$3.200, 3.206$]$\gevcc$.
Figure~\ref{fig:fitting2d} shows the projections of
the two-dimensional fit results
and Table~\ref{tbl:num_fit} lists the numerical results. The
$\chi^2$ per degree of freedom corresponding to the fit is 318/349. The statistical
significance of $\jpsi\to3\gm$ ($\jpsi\to\gm(\gm\gm)_{\etac}$) is
8.3$\sigma$ (4.1$\sigma$), as determined by the ratio of the maximum
likelihood value and the likelihood value for a fit under the null
hypothesis. When the systematic uncertainties are included, the
significance becomes 7.3$\sigma$ (3.7$\sigma$).  The branching
fraction is calculated using
\begin{equation}
\mathcal{B}=\frac{n_{\unit{obs}}}{N_{\psip}\times\br{\psip\to\pi^+\pi^-\jpsi}\times
\varepsilon}
\label{eq:bf}
\end{equation}
 where $n_{\unit{obs}}$ is the observed number of events,
$N_{\psip}$ is the number of $\psip$ events~\cite{Ablikim:2012pj}, and
$\varepsilon$ is the detection efficiency.  The branching fraction for $\psip\to\pi^+\pi^-\jpsi$ is taken from the PDG~\cite{PDG2012}. Simulation of direct
$\jpsi\to3\gm$ decay assumes the lowest order matrix element is similar
to the decay of ortho-positronium to three
photons~\cite{PhysRevLett.76.4903}.

%%%%%%%%%%%%%%%%%%%%%%%%%%%%%%%%%%%%%%%%%%%%%%%%%%%
%%%%%%%%%%%%%%%%%%%%%%%%%%%%%%%%%%%%%%%%%%%%%%%%

%%%%%%%%%%%%%%%%%%%%%%%%%%%%%%%%%%%%%%%%%%%%%%
%%%%%%%%%%%%%%%%%%%%%%%%%%%%%%%%%%%%%%%%%%%%%
%CN - what about the blue histogram?

Sources of systematic uncertainty in the measurement are listed in Table~\ref{tab:syst_err}.
For the process $\jpsi\to3\gm$, there is no explicit theoretical input for the matrix element.
The signal model used in the simulation determines the uncertainty in estimating the detection efficiency.
In the kinematic phase space in the Dalitz-like plot of Fig.~\ref{fig:proj}(e), the detection efficiency, $\varepsilon$,
is formulated as
\begin{equation}
\varepsilon
=\sum_{i,j}\frac{N^{ij}}{\sum_{i,j}N^{ij}}\varepsilon^{ij}
=\frac{\sum_{i,j}n^{ij}}{\sum_{i,j}\frac{n^{ij}}{\varepsilon^{ij}}}
\label{eq:eff_2D}
\end{equation}
where $N^{ij}=\frac{n^{ij}}{\varepsilon^{ij}}$ is the number of
acceptance-corrected signals, $n^{ij}$ is the number of
observed signals, and $\varepsilon^{ij}$ is the detection efficiency in kinematic bin ($i,j$).
MC studies show that $\varepsilon^{ij}$ ranges from 34.0\% to 39.1\%.
Given a sufficient yield, Eq.~\eqref{eq:eff_2D} would provide a realistic unbiased $\varepsilon$
from the weighted sum of $\varepsilon^{ij}$.
However, this is not applicable in this work due to the low statistics of the signal yield.
With a reasonable assumption that signal yields are continuously distributed over the full phase space in Fig.~\ref{fig:proj}(d), the maximum relative change of $\varepsilon^{ij}$, 15\%, is taken as the systematic uncertainty.
For the case of $\jpsi\to\gm\etac$, its decay mechanism is well understood and the corresponding uncertainty is negligible.

The invariant mass of the $\etac$ in the $\jpsi\to\gm\etac$ decay is
assumed to have a relativistic Breit-Wigner distribution, weighted by
a factor of $E^{*3}_\gm$ multiplied by a damping factor
$e^{-E^{*2}_\gm/8\beta^2}$, with
$\beta=(65.0\pm2.5)\mev$~\cite{:2008fb}. Here, $E^*_\gm$ is the energy
of the radiated photon in the $J/\psi$ rest frame. An alternative
parametrization of the damping factor used by
KEDR~\cite{Anashin:2010dh} changes the measurement by 1\%, which is taken as the systematic uncertainty in the $\etac$ line shape.
In addition, variations of the $\etac$ width in the range 22.7--32.7\,MeV affect the measurement
of $\br{\jpsi\to\gm\etac,\etac\to\gm\gm}$ by 5\%.

%%%%%%%%%%%%%%%%%%%%%%%%%%%%%%%%%%%%%%%%%%%%%
\begin{table}[t]
  \centering
    \caption{Estimated numbers of events for the backgrounds shown in Fig.~\ref{fig:fitting2d}.}\label{tbl:num_bkg}
  \begin{tabular}{l|c|c}
    \hline
     Channels            &   Survival rate (\%)           & Number of events \\
    \hline
    $\jpsi\to\gm\pi^0$   &   0.45  & $5.6\pm0.5$  \\
    $\jpsi\to\gm\eta$    &   0.47  & $72.9\pm2.4$ \\
    $\jpsi\to\gm\eta'$   &   0.44  & $18.2\pm0.8$ \\
    Non-$\jpsi$ decays   &         & $20\pm4.5$ \\
    \hline
  \end{tabular}
\end{table}
%%%%%%%%%%%%%%%%%%%%%%%%%%%%%%%%%%%%%%%%%%%%%%%

%%%%%%%%%%%%%%%%%%%%%%%%%%%%%%%%%%%%%%%%%%%%%
\begin{table}[t]
  \centering
    \caption{The detection efficiency $\varepsilon$, signal yields,
    estimated significance and measured branching fractions,
    with their uncertainties, for the two decay modes. The first set of uncertainties are
    statistical and the second are systematic. Values of the significance
    outside the parenthesis are statistical only and those within the
    parenthesis include systematic effects.}\label{tbl:num_fit}
  \begin{tabular}{l|cc}
    \hline
     Mode            &   $\jpsi\to3\gm$            & $\jpsi\to\gm\etac,\etac\to\gm\gm$ \\
    \hline
    $\varepsilon$ (\%)          &   $27.9\pm0.1$          & $20.7\pm0.2$                              \\
    Yield                 &   $113.4\pm18.1$                & $33.2\pm9.5$                        \\
    Significance           &   $8.3(7.3)\sigma$                   & $4.1(3.7)\sigma$                         \\
    $\mathcal{B}$($\times 10^{-6}$)  &   $11.3\pm1.8\pm2.0$                  & $4.5\pm1.2\pm0.6$                         \\
    \hline
  \end{tabular}
\end{table}
%%%%%%%%%%%%%%%%%%%%%%%%%%%%%%%%%%%%%%%%%%%%%%%

The systematic uncertainty due to possible bias in modelling the
detector resolutions is evaluated by performing a two-dimensional fit of the
$\chisq{4C}$ and $M(\gm\gm)_{\unit{lg}}$ distributions with
MC shapes smeared by an asymmetric Gaussian function. The function
parameters are determined by comparing a $\jpsi\to\gm\eta$, $\eta\to\gm\gm$
control data sample to a corresponding simulated sample.  This function
serves to adjust the detector resolution in the MC simulation to that seen in the data.
Inclusion of this resolution function changes the numerical results
by 3\% for $\br{\jpsi\to3\gm}$ and 9\% for $\br{\jpsi\to\gm\etac, \etac\to\gm\gm}$.

\begin{figure}[t]
    \centering
    \includegraphics[width=0.9\linewidth]{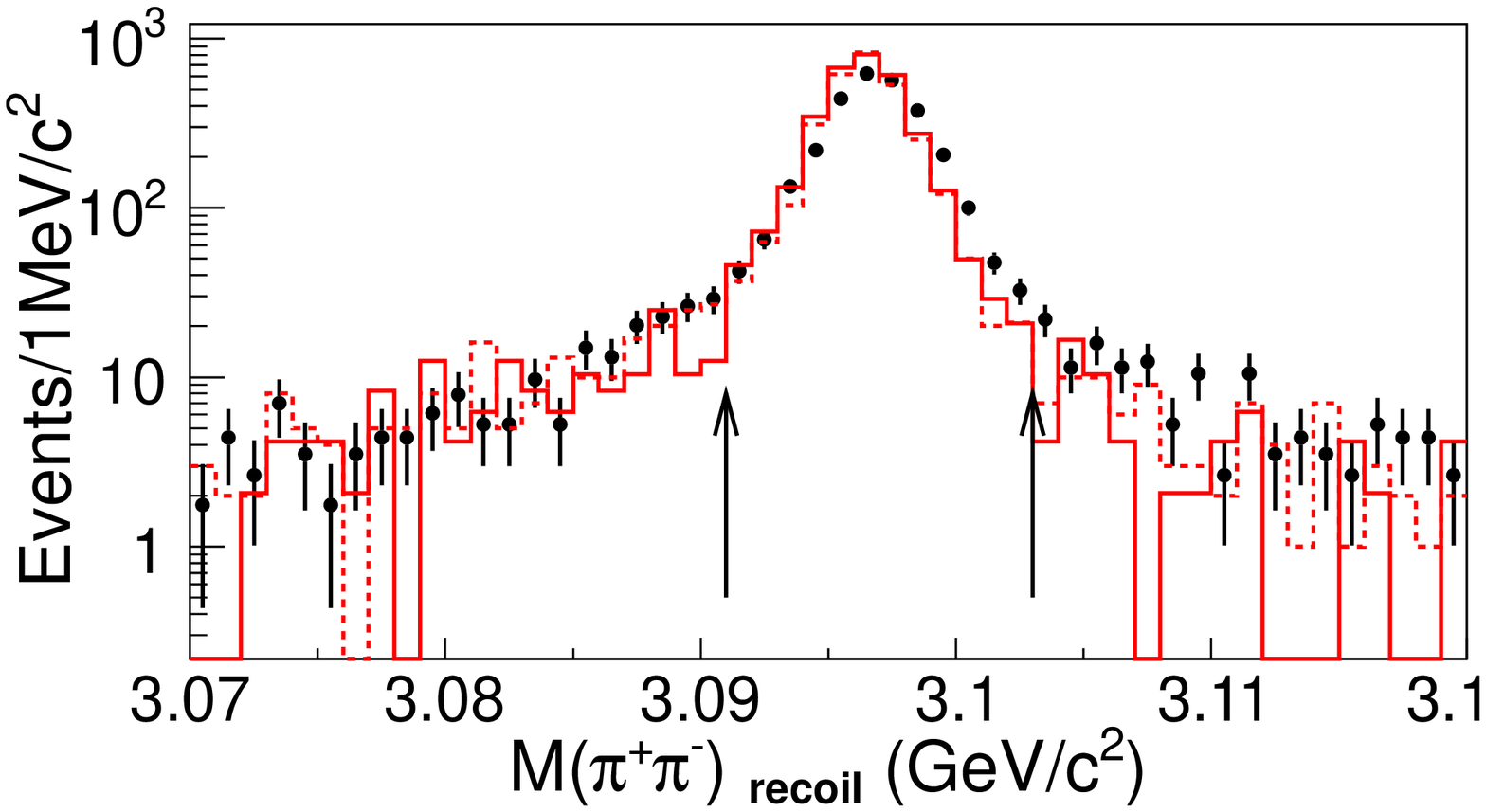}
    \vspace{-0.25cm}
    \caption{ The $\pi^+\pi^-$ recoil mass spectrum $M(\pi^+\pi^-)_{\unit{recoil}}$
    from the control channel $\psip\to\pi^+\pi^-\jpsi,\jpsi\to\gm(\gm\gm)_{\eta}$
    for data (points with error bars) and MC simulation (dashed histogram).
    The event selection is the same as for $\jpsi\to3\gm$ but with the requirement
    that the $\gm\gm$ mass $M(\gm\gm)$ has to lie within [0.5, 0.6]\,$\gevcc$.
    For comparison, a simulation of $\jpsi\to\gm(\gm\gm)_{\eta}$ is shown (solid histogram) with the area scaled to that of the full $\psip$ decay chain simulation. The arrows indicate the signal region for selection of $\jpsi$ events.}
    \vspace{-0.25cm}
    \label{fig:recoilmass}
\end{figure}
%%%%%%%%%%%%%%%%%%%%%%%%%%%%%%%%%%%%%%%%%%%%%%%%%%%%%%%%%%%%%%%

Figure~\ref{fig:recoilmass} compares $M(\pi^+\pi^-)_{\unit{recoil}}$ distributions in data and MC simulation for $\psip$ inclusive decays, based on the $\psip\to\pi^+\pi^-\jpsi,\jpsi\to\gm(\gm\gm)_{\eta}$ control samples. It also shows the distribution for a dedicated MC simulation of the process $\jpsi\to\gm(\gm\gm)_{\eta}$.
As Fig.~\ref{fig:recoilmass} shows, there is a slight discrepancy between data and MC simulation in the position of the peak in the $M(\pi^+\pi^-)_{\unit{recoil}}$ spectrum. This discrepancy is due to
the tracking simulation of low momentum pions.
Since the $\jpsi$ mass window is sufficiently broad to cover the peak region in both data and MC simulation, the efficiency of the mass window requirement should not be significantly affected. The relevant systematic uncertainty is studied with a $\jpsi\to\gm\eta$, $\eta\to\gm\gm$ control sample. Using different mass window regions give a maximum change of 4\% in
$\br{\jpsi\to\gm\eta}$; this is therefore taken as the systematic uncertainty.

The uncertainty in the expected number of background
events from $\jpsi\to\gm\pi^0$($\eta$, $\etap$) is evaluated by varying
their branching fractions by one standard deviation~\cite{PDG2012}.
The maximum changes in the results are 0.5\% for $\br{\jpsi\to3\gm}$ and 5\% for $\br{\jpsi\to\gm\etac,\etac\to\gm\gm}$.

%%%%%%%%%%%%%%%%%%%%%%%%%%%%%%%%%%%%%%%%%%%%%%
\begin{figure}[t]
\centering
    \footnotesize
%    \vspace{-0.3cm}
    \includegraphics[width=0.8\linewidth]{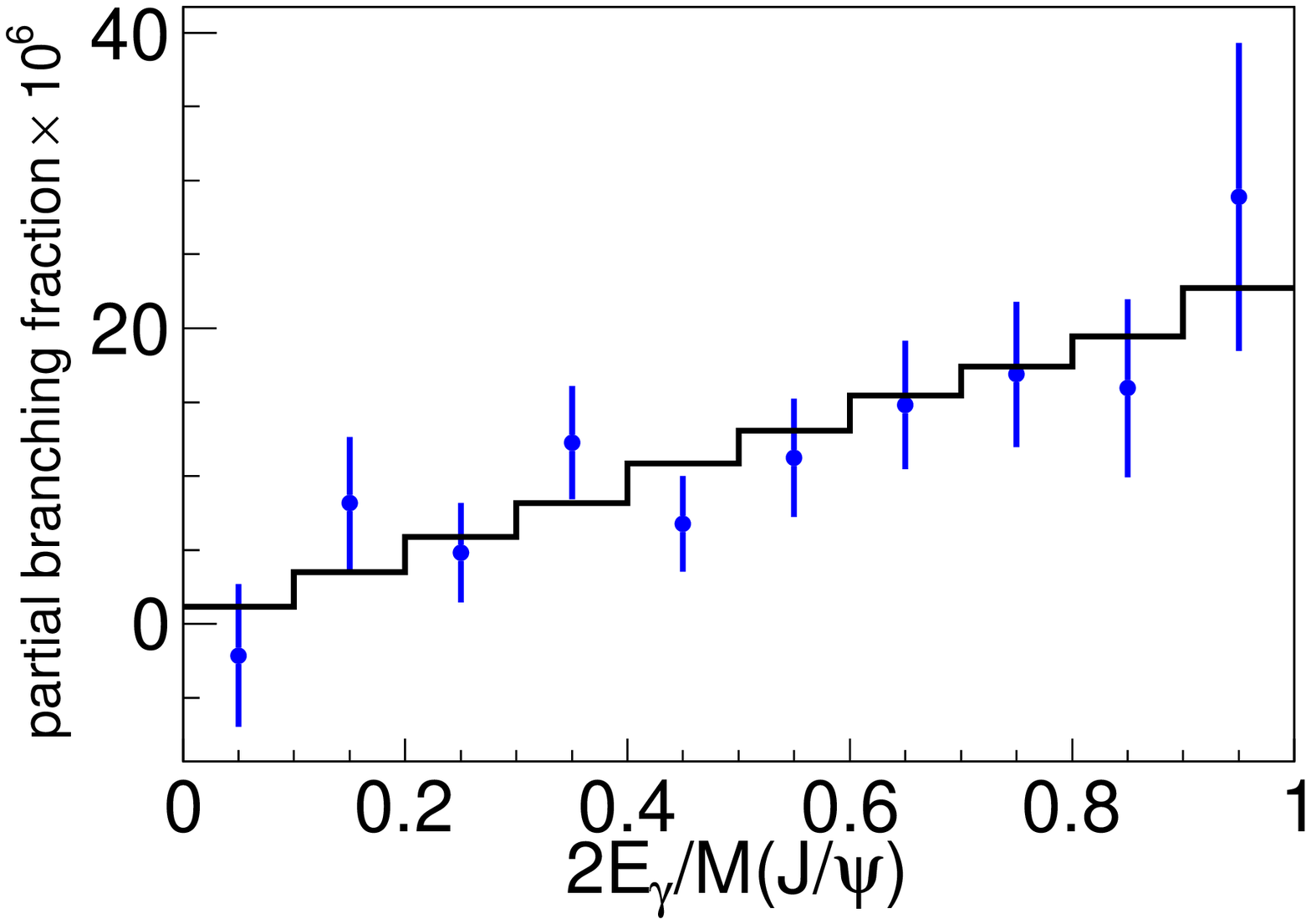}
    \caption{The energy spectrum of $\jpsi\to3\gm$
        inclusive photons in the $\jpsi$ rest frame . The points with error bars
    represent the partial branching fractions as a function of the ratio
    $2E_{\gamma}/M_{\jpsi}$ measured in data. Here, $E_{\gamma}$ is the
    photon energy and $M_{\jpsi}$ is the $\jpsi$ mass. The solid line
    shows the theoretical calculation according to the
    ortho-positronium decay formula~\cite{PhysRevLett.76.4903}.  }
    \vspace{-0.25cm}
    \label{fig:spectrum}
\end{figure}
%%%%%%%%%%%%%%%%%%%%%%%%%%%%%%%%%%%%%%%%%%%%%

\begin{table}[t]
  \begin{center}
  \caption{ Summary of the relative systematic uncertainties. $\mathcal{B}_{3\gm}$ and   $\mathcal{B}_{\gm\etac}$ stand for the measurements of branching fractions $\br{\jpsi\to3\gm}$ and $\br{\jpsi\to\gm\etac,\etac\to\gm\gm}$, respectively. A dash (--) means
  the uncertainty is negligible.}
  \label{tab:syst_err}
    \begin{tabular}{lcc}
      \hline
      \hline
      \multirow{2}{*}{Source } & \multicolumn{2}{c}{\centering{Uncertainties (\%)}}
      \\ \cline{2-3}
      &  $\mathcal{B}_{3\gm}$ &   $\mathcal{B}_{\gm\etac}$ \\  \hline
      Signal model                    & 15                     &-- \\
      $\etac$ width                  & --                        & 5 \\
      $\etac$ line shape               & 1                      & 1 \\
      Resolution                     & 3                       & 9\\
      $M(\pi^+\pi^-)_{\unit{recoil}}$ window         & 4                   & 4 \\
      $\pi^0, \eta, \eta'$ rejection    & 0.5                      &5 \\
      PWA model                      & 2                      & 2 \\
      Photon detection                  & 3                      &3\\
      Tracking                          & 2                      &2\\
      Number of good photons            & 0.5                      & 0.5 \\
      Kinematic fit and $\chisq{4C}$ requirement                  & 2                      & 2\\
      Fitting                           & 5                      & 5\\
      Number of $\psip$        & 0.8                      & 0.8 \\
       $\br{\psip\to\pi^+\pi^-\jpsi}$ &1.2       &1.2  \\ \hline
      Total            & 18&  14   \\
      \hline \hline
    \end{tabular}
    \end{center}
\end{table}

It has been verified that the $\chisq{4C}$ distribution of the
$\gm\pi^0\pi^0$ final states does not depend on the components of the
intermediate processes involved; in this case, these are mainly the $f_J$
states~\cite{Adams:2008ab}. Since the $M(\gm\gm)_{\unit{lg}}$ mass
distribution does depend on the components of the intermediate structures, however,
it is important to obtain a good understanding of the primary
components using PWA. Information about the amplitudes in
$\jpsi\to\gm\pi^0\pi^0$ from the previous BESII
analysis~\cite{Ablikim:2006db} is also used in the simulation as
an additional check; the relative change of 2\% in the results is taken as
the systematic uncertainty due to the PWA model.

The photon detection efficiency is studied with different control
samples, such as radiative Bhabha and $\psip\to\pi^+\pi^-
\jpsi$, $\jpsi\rightarrow \rho^0\pi^0$ events~\cite{Collaboration:2010rc}.
A systematic uncertainty of 1\% is assigned for each photon over the kinematic region covered in this work, so
a total of 3\% is assigned for the three photons in the final states studied.
The MDC tracking efficiency is studied using selected samples of
$\jpsi\to\rho\pi$ and $\psip\to\pi^+\pi^-\jpsi$,
$\jpsi\to\pi^+\pi^-p\overline{p}$ events~\cite{Ablikim:2012hi}.
The disagreement between data and MC simulation is within 1\% for each pion, so 2\% is assigned as the total systematic uncertainty for the two pions.
Samples of $\jpsi\to\gm\eta$, $\eta\to\gm\gm$ events are selected to study uncertainties arising from requirements on the number of photon
candidates and the $\chisq{4C}$ requirement, which are given as 0.5\% and 2\%, respectively.
The uncertainty due to the fitting is estimated to be 5\% by changing the fitting range and the bin
width.

The uncertainty in determining the number of $\psip$ events is 0.8\%~\cite{Ablikim:2012pj}. The uncertainty in $\br{\psip\to\pi^+\pi^-\jpsi}$ is taken to be 1.2\%, as quoted by the PDG~\cite{PDG2012}.

The energy spectrum of inclusive photons in $\jpsi\to3\gm$ provides
information on the internal structure of the
$\jpsi$~\cite{Voloshin:2007dx}. An inclusive photon is defined as any
one of the three photons in the final state. Partial branching
fractions are measured as a function of inclusive photon energy
$E_\gm$ in the $\jpsi$ rest frame. Figure~\ref{fig:spectrum} shows the model-independent
photon energy distribution as measured for all three photons from $\jpsi\to3\gm$,
where the error bars are combinations of the statistical and
systematic uncertainties. The distribution agrees well with the
theoretical calculation adapted from the ortho-positronium decay model.
However, the experimental uncertainties are still rather large.

In conclusion, the $\jpsi$ decays to three photons are studied using
$\psip\to\pi^+\pi^-\jpsi$ decays at BESIII. The
direct decay of $\jpsi\to3\gm$ is measured to be
$\br{\jpsi\to3\gm}=(11.3\pm1.8\pm2.0)\times 10^{-6}$, which
is consistent with the result from CLEO. Combining the results of the two experiments gives
$\br{\jpsi\to3\gm}=(11.6\pm2.2)\times 10^{-6}$.
With the input of $\br{\jpsi\to e^+e^-}$ from the PDG~\cite{PDG2012}, $\mathcal{R}$ is then determined
to be $(1.95\pm0.37)\times10^{-4}$. This is clearly incompatible with the
calculation in Eq.~\eqref{eq:jpsi_ratio}, which indicates that further improvements of the
QCD radiative and relativistic corrections are needed. A study in Ref.~\cite{Feng:2012by}
reveals that the discrepancy can be largely remedied by introducing the joint perturbative
and relativistic corrections.

The energy spectrum of inclusive
photons in $\jpsi\to3\gm$ is also measured.
Evidence of the $\etac\to\gm\gm$ decay is reported, and the product branching fraction of $\jpsi\to\gm\etac$ and $\etac\to\gm\gm$ is determined to be
$\br{\jpsi\to\gm\etac,\etac\to\gm\gm}=(4.5\pm1.2\pm0.6)\times10^{-6}$.
This result is consistent with the theoretical
prediction~\cite{Kwong:1987ak} and the CLEO result~\cite{Adams:2008ab}.
When combined with the input of
$\br{\jpsi\to\gm\etac}=(1.7\pm0.4)\times10^{-2}$ from the
PDG~\cite{PDG2012}, we obtain
$\br{\etac\to\gm\gm}=(2.6\pm0.7\pm0.7)\times10^{-4}$, which
agrees with the result from two-photon fusion~\cite{PDG2012}.

%%%%%%%%%%%%%%%%%%%%%%%%%%%%%%%%%%%%%%%%%%%%%%%%%%%%%%%%%%%%%%%%
%%%%%    acknowledgments       Part                %%%%%%%%%%%%%
%%%%%%%%%%%%%%%%%%%%%%%%%%%%%%%%%%%%%%%%%%%%%%%%%%%%%%%%%%%%%%%%
The BESIII collaboration thanks the staff of BEPCII and the IHEP computing
center for their hard work. This work is supported in part by the
Ministry of Science and Technology of China under Contract No.
2009CB825200; National Natural Science Foundation of China (NSFC)
under Contracts Nos.
10625524, 10821063, 10825524, 10835001, 10935007, 10905091,
11125525; Joint Funds of the National Natural
Science Foundation of China under Contracts Nos. 11079008, 11179007;
the Chinese Academy of Sciences (CAS) Large-Scale Scientific Facility
Program; CAS under Contracts Nos. KJCX2-YW-N29, KJCX2-YW-N45; 100
Talents Program of CAS; Istituto Nazionale di Fisica Nucleare, Italy;
U. S. Department of Energy under Contracts Nos. DE-FG02-04ER41291,
DE-FG02-91ER40682, DE-FG02-94ER40823; U.S. National Science
Foundation; University of Groningen (RuG) and the Helmholtzzentrum
fuer Schwerionenforschung GmbH (GSI), Darmstadt; WCU Program of
National Research Foundation of Korea under Contract No.
R32-2008-000-10155-0.

%%%%%%%%%%%%%%%%%%%%%%%%%%%%%%%%%%%%%%%%%%%%%%%%%%%%%%%%%%%%%%%%
%%%%%    bibliographies       Part                %%%%%%%%%%%%%
%%%%%%%%%%%%%%%%%%%%%%%%%%%%%%%%%%%%%%%%%%%%%%%%%%%%%%%%%%%%%%%%

\end{document}